\documentclass{iaus}
\usepackage{graphicx}

\def\+/-{{\pm}}
\def\=={{\equiv}}
\def\Beq{B_{eq}}
\def\Rstar{R_{\ast}}

\def\vinf{v_\infty}

\title[Spin-down of Massive Stars] 
{Angular Momentum Loss and Stellar spin-down in Magnetic Massive Stars}

\author[ud-Doula et al.]   
{Asif ud-Doula$^1$%
  \thanks{email address: uddoula@morrisville.edu},
 Stanley P. Owocki$^2$ \break \and Richard H.D. Townsend$^3$}

\affiliation{$^1$Department of Physics, SUNY-Morrisville State College,
Morrisville, NY 13408, USA \\[\affilskip]
$^2$Department of Physics and Astronomy, University of
Delaware, \break Newark, DE 19716, USA \\[\affilskip]
$^3$Department of Physics, University of
Wisconsin-Madison, \break Madison, WI 53706, USA }

\pubyear{2009}
\volume{259}  
\pagerange{100--100}
\date{"Dec 1, 2008"  and in revised form ??}
 \jname{Cosmic Magnetic Fields: From Planets,
to Stars and Galaxies} \editors{K.G. Strassmeier, A.G. Kosovichev \&
J.E. Beckman, eds.}
\begin{document}

\maketitle

\begin{abstract}
We examine the angular momentum loss and associated rotational
spin-down for magnetic hot stars with a line-driven stellar wind
and a rotation-aligned dipole magnetic field. Our analysis here
is based on our previous 2-D numerical MHD simulation study
that examines the interplay among wind, field, and rotation as
a function of two dimensionless parameters, W(=Vrot/Vorb) and
'wind magnetic confinement', $\eta_\ast$ defined below.
We compare and contrast the 2-D, time variable angular momentum
loss of this dipole model of a hot-star wind with the classical
1-D steady-state analysis by Weber and Davis (WD), who used an
idealized monopole field to model the angular momentum loss in the
solar wind. Despite the differences, we find that the total angular
momentum loss averaged over both solid angle and time
follows closely the general WD scaling $\dot {J} \sim \dot {M} \Omega R_A^2$.
The key distinction is that for a dipole field Alfv\`en radius $R_A$ is
significantly smaller than for the monopole field WD used in their
analyses. This leads to a slower stellar spin-down for the dipole
field with typical spin-down times of order 1~Myr
for several known magnetic massive stars.
\end{abstract}

\firstsection 
\section{Introduction}

An outflowing wind carries away angular momentum and
thus spins down the stellar rotation. Winds with magnetic
fields exert a braking torque that is significantly larger than
for non-magnetic cases, due to the larger lever arm of magnetic
field lines that extend outward from the stellar surface.
A seminal analysis of this process was carried out by
Weber \& Davis (WD, 1967), who modelled the angular momentum loss of the solar
wind for the idealized case of a simple monopole magnetic
field from the solar surface. In terms of the surface angular
velocity $\Omega$ and wind mass loss rate $\dot M$, Weber \& Davis
concluded that the total
angular momentum loss rate scales as:
\begin {equation}
\dot J= \frac {2}{3} \dot M \Omega R_A^2 \, ,
\label{jdot-eqn}
\end {equation}
with $R_A$ the Alfv\'en radius,defined by where the radial
components of the field and flow have equal energy density.
However, WD did not provide a prescription for computing
such an Alfv\'en radius even for the idealized monopole field geometry.

For any radius $r$, the energy density ratio between radial field and flow
is given by
\begin{equation}
\eta(r) \equiv \frac{B_{r}^{2}/8\pi}{\rho v_{r}^{2}/2}
\, .
\label{eta-eqn}
\end{equation}
The Alfv\'{e}n radius is then defined implicitly by
$\eta(R_{A})\equiv 1$.
We can derive approximate {\em explicit} expressions in terms
of fixed  values for the equatorial field strength $B_{eq}$ at the
surface radius $R_{\ast}$, and for the wind mass loss rate ${\dot M}$ and
terminal flow speed $v_\infty$.
Specifically, following ud-Doula \& Owocki (paper 1, 2002) and
ud-Doula et al. (paper 2, 2008a), if we define here a wind
{\em magnetic confinement parameter},
\begin{equation}
\eta_{\ast} \equiv \frac {\Beq^2 \, \Rstar^2} {\dot{M} \, \vinf}
\, ,
\label{esdef}
\end{equation}
then we can write the energy density ratio in the form
\begin{equation}
\eta(r) =  \eta_{\ast}
\,
\left [ \frac{r}{\Rstar} \right ]^{2-2q}
\, \frac{\vinf}{v_{r}(r)}
\end{equation}
where $q$ is the power-law exponent for radial decline of
the assumed magnetic field
and terminal speed $v_\infty$. It turns out that
in the strong magnetic
confinement limit $\eta_{\ast} \gg 1$, for a monopole
field $R_{A} \sim \sqrt{\eta_{\ast}}$ whereas for a dipole field
the scaling is significantly weaker $R_{A} \sim {\eta_{\ast}}^{1/4}$.

\section{Key Results}
In our recent paper 3 (\cite{udDOwo2008b}), which this poster
summarizes,  we examine the wind
magnetic spin-down of massive stars with a rotation-aligned
dipole field based on previous magnetohydrodynamic (MHD)
simulation parameter study presented in paper 2.
Despite key differences, we find that the total angular
momentum loss from these massive stars follow
the general WD scaling \ref{jdot-eqn}.
However,  because for dipole fields the Alfv\'en radius
has a stronger field scaling than for the idealized
monopole case, the net stellar spin-down time is
also significantly longer. Our numerical simulations show that
this spin-down time can be expressed as:
\begin{equation}
\frac
{\tau_{spin}}
{\tau_{mass}}
\approx
\frac
{\frac{3}{2} k}
{\left [ 0.29 + (\eta_{\ast} + 0.25)^{1/4} \right ]^{2}}
\, .
\label{tauspin-dwd}
\end{equation}
where $\tau_{mass}$ is the characteristic mass loss rate ($=M_\ast/\dot M$),
$k$ is the moment of inertia constant with a typical value $k \approx 0.1$.
This leads to typical spin-down time of  $\sim$1 Myr for several
know massive stars.
Details of our full calculations can be found in \cite{udDOwo2008b}.

\begin{acknowledgments}
This work was supported by NASA Grants Chandra/TM7-8002X and NNX08AY04G.
\end{acknowledgments}

\end{document}